\RequirePackage{fix-cm}
\documentclass[twocolumn,epjc3]{svjour3}
\smartqed  
\RequirePackage{graphicx}
 \RequirePackage{mathptmx}      
\RequirePackage{latexsym}
\RequirePackage[numbers,sort&compress]{natbib}
\RequirePackage[colorlinks,citecolor=blue,urlcolor=blue,linkcolor=blue]{hyperref}

\usepackage[T1]{fontenc}
\usepackage{latexsym}
\usepackage{amsfonts}
\usepackage[usenames,dvipsnames]{xcolor}
\usepackage{graphicx,amssymb,amsmath,epsfig}
\usepackage[english]{babel}

\usepackage{bm}
\usepackage{slashed}
\usepackage{calligra}
\usepackage{dsfont}
\usepackage{braket}
\usepackage{slashed}

\newcommand{\be}{\begin{equation}}
\newcommand{\ee}{\end{equation}}
\newcommand{\br}{\begin{eqnarray}}
\newcommand{\er}{\end{eqnarray}}

\def\({\left(}
\def\){\right)}

\newcommand{\da}{\dagger}
\newcommand{\g}{\gamma}

\newcommand{\pa}{\partial}
\newcommand{\s}{\sigma}
\def \th{\theta}
\newcommand{\pv}{{\bf p}}

\journalname{Eur. Phys. J. C}
\begin{document}
\sloppy

\title{Spin-1/2 ``bosons'' with mass dimension 3/2 and fermions with mass dimension 1 cannot represent physical particle states}


\author{A. R. Aguirre\thanksref{addr1,e1}
       \and
        M. M. Chaichian \thanksref{addr2,e2}
        \and
        B. A. Couto e Silva\thanksref{addr3,e3}
        \and
        B. L. S\'anchez-Vega\thanksref{addr3,e4}
       }

\thankstext{e1}{e-mail: alexis.roaaguirre@unifei.edu.br}
\thankstext{e2}{e-mail: masud.chaichian@helsinki.fi}
\thankstext{e3}{e-mail: brunoaces@ufmg.br}
\thankstext{e4}{e-mail: bruce@fisica.ufmg.br}


\institute{Instituto de F\'isica e Qu\'imica, Universidade Federal de Itajub\'a, Av. BPS 1303, Itajub\'a -- MG, CEP 37500-903, Brazil. \label{addr1}
\and
    Department of Physics, University of Helsinki, P.O. Box 64, 00014 Helsinki, Finland. \label{addr2}
         \and
           Departamento de F\'isica, UFMG, Belo Horizonte, MG 31270-901, Brazil. \label{addr3}
           }

\date{Received: date / Accepted: date}

\maketitle

\begin{abstract}
{We delve into the first principles of quantum field theory to prove that the so-called spin-1/2 ``bosons'' and the fermions with mass dimension 1, including ELKO, cannot represent physical particle states with spin $1/2$. Specifically, we first demonstrate that both aforementioned fields are not invariant under rotational symmetry, which implies that the particles created for these fields are not eigenstates of the spin operator in the $(\frac{1}{2},0)\oplus(0,\frac{1}{2})$ representation of the Lorentz group, nor is it possible to construct a Hamiltonian density scalar under the rotational group from them. Furthermore, following Weinberg's approach to local causal fields, we prove that regardless of any discrete symmetry or adjoint structure, the relativistic fields in the $(\frac{1}{2}, 0) \oplus (0,\frac{1}{2})$ representation satisfy the Fermi-Dirac statistics in complete agreement with the well-established spin-statistics theorem and experimental results.}

\end{abstract}


\section{Introduction}
\label{intro}

Wigner's pioneering work established the concept of particle in physics \cite {WIG, WIG2}. It has been demonstrated, in a well-posed mathematical formulation, that a particle is nothing more than an irreducible representation of the Poincar\'e group. Wigner first considered the proper orthochronous Lorentz group \cite{WIG}, which is the subgroup of the Lorentz group that preserves orientation and the direction of time. Then he achieved a general treatment of the states of particles by including the reflections \cite{WIG2}.

The foundations of Quantum Field Theory (QFT) dictate that quantum fields are the result of engaging well defined one-particle states into interactions regulated through Lorentz invariance \cite{weinbergfeynman}. Furthermore, fundamental rules such as the principles of cluster decomposition \cite{wic} and causality do form the theoretical setting on which a quantum field emerges \cite{WEI,Duncan}.

In recent years, however, several authors have explored alternative ways to introduce new quantum fields with spin $1/2$. Such proposals have been motivated by the need to explain the unsolved problem of dark matter in the Universe.
The proposed fields, in particular the so called ELKO fields (an acronym from the German {\it Eigenspinor des Ladungskonjugation Operators}) \cite{Grumiller2005,PhysRevD.72.067701} which are based on the spinor classification proposed by P. Lounesto \cite{lounesto}, have attracted considerable attention due to their unusual characteristics.
Fermions with mass dimension 1 have been introduced in Refs.~\cite{Grumiller2005,PhysRevD.72.067701,mdobook,newfermions} and widely explored in cosmology, mathematical physics and phenomenology \cite{Boehmer1,Boehmer2,Boehmer3,Boehmer4,Boehmer5,Fabbri,Sadjadi,Basak,daRocha:2007pz,daRocha:2011yr,Fabbri2,Agarwal:2014oaa,Alves:2014kta,HoffdaSilva:2016ffx,CYL,nogo,ahluwa1,Rogerio2017,elkostates}.
These fields have some properties that differ from the fermionic fields of the Standard Model of elementary particles, where fermions have the mass dimension $3/2$, and for that reason the new fields were considered as candidates for dark matter. For a more recent review on this subject we refer to \cite{report} and references therein.
On the other hand, a new class of spin-1/2 fields was proposed in Refs. \cite{dharamnewbosons,Ahluwalia:2021vfu,Ahluwalia:2019ujt} that apparently obey the Bose-Einstein statistics, which differs from what is well-established by the spin-statistics theorem.

The very well-known spin-statistics theorem, in a (3+1)-dimensional spacetime, states that particles with half-integer spin follow the Fermi-Dirac statistics, that is, they are fermions. On the other hand, particles with integer spin follow the Bose-Einstein statistics, that is, they are bosons. This spin-statistics connection is a fundamental principle of physics with dramatic physical consequences even in a non-relativistic context, e.g. the effects of Pauli's exclusion principle on electrons in atoms and metals, the effects of Bose-Einstein condensation in superfluids, superconductors and lasers. The spin-statistics connection has been demonstrated several times using different general assumptions. See \cite{Curceanu} for a complete review of the literature on the spin-statistics theorem. Among the most common assumptions we can mention microcausality, the positive semidefinite Hamiltonian, locality of fields and interactions and positive norms for states. Among the most remarkable proofs, we can mention Pauli's demonstration \cite{Pauli1936} based on classical microcausality \cite{Massimi} and the positive semidefinite Hamiltonian conditions \cite{Pauli1940}. In addition, the spin-statistic theorem was also proven using Axiomatic quantum field theory \cite{Wightman}, i.e. without referring to any Lagrangian or Hamiltonian. A more modern version of the proof is due to S. Weinberg \cite{weinbergfeynman, WEI}.  Roughly speaking, Weinberg's approach was motivated by the construction of a general theory with a Lorentz invariant $S$-matrix. The main advantage of this approach is that it does not assume any particular Lagrangian or equation of motion, i.e. any particular dynamics, to deduce the connection between spin and statistics.

From an experimental point of view, the spin-statistics theorem has been tested, with a high level of precision, for different types of particles. For particles with half-integer spin, one of the pioneering experimental tests of the Pauli exclusion principle, which is a consequence of the spin-statistics theorem, was conducted by Goldhaber and Scharff-Goldhaber in 1948 \cite{Goldhaber}. Other experimental searches for violation of the Pauli exclusion principle for electrons and nucleons were done in \cite{Reines1974954,Amado19801338,Deilamian19954787,Borexino:2004hfc}. All of them obtained negative results. Among the most recent experiments, we can mention the VIP-2 (Violation of the Pauli Principle), which is an atomic physics experiment, located in the underground laboratory of Gran Sasso, where the possible violation of the Pauli exclusion principle for electrons is studied. This experimental collaboration performed a statistical analysis of approximately six months of data collection between December 2019 and May 2020, constraining  $\beta^2/2$ (with $\beta$ the span parameter connecting prohibited states to permitted states) to less than $6.8 \times 10^{-42}$ at the $90\%$ C.L. \cite {sym14050893}. On the other hand, one of the experimental tests of the Bose-Einstein (BE) statistics and, consequently, of the spin-statistics theorem for photons interacting with atoms was carried out at Lawrence Berkeley National Laboratory in California, USA. This experimental collaboration constrained the rate of transitions that violate the statistics, as a fraction $\nu$ of an equivalent transition rate allowed by the statistics, to $\nu <4.0 \times 10^{-11}$ at $90\%$ C.L. \cite {PhysRevLett.104.253604}.  In conclusion, according to the experiments, the spin-statistics theorem remains valid.

Therefore, since the spin-statistics theorem is established so well both theoretically and experimentally, and has been revisited by many authors for over 80 years, it is very important to scrutinize the theoretical proposal of \cite{dharamnewbosons,Ahluwalia:2021vfu,Ahluwalia:2019ujt} in order to find out which assumptions of the spin-statistics theorem are violated for the so-called spin-1/2 ``bosons''.

In order to answer this question, we organize this article in the following way. In section \ref{sec2}, we follow the approach of local causal fields \cite{WEI}, and use the general principle of Lorentz invariance to find the fundamental conditions that any spin-1/2 field, in the $(\frac{1}{2},0)\oplus(0,\frac{1}{2})$ representation space, must satisfy. In section \ref{sec3}, we consider the quantum field proposed in \cite{dharamnewbosons,Ahluwalia:2021vfu,Ahluwalia:2019ujt} and show that it does not satisfy the rotational symmetry conditions for spin-1/2 fields, implying that it can not be used to describe a physical particle state with spin $1/2$. In section \ref{sec4}, we apply the same procedure to show that the fermion fields with mass dimension 1 proposed in \cite{Grumiller2005,PhysRevD.72.067701,mdobook,newfermions} also fail to satisfy the conditions for rotational symmetry and, therefore, it cannot represent a physical state with spin $1/2$. In addition, based on arguments from the representation theory of groups, we provide the reason why such proposed fields do not satisfy the rotational symmetry conditions.
In section \ref{sec5}, we use a generalized version of the Weinberg approach, which is based on the  Lorentz invariance of the $S$-matrix and the cluster decomposition principles, in order to show that any spin $1/2$ quantum field must satisfy anti-commutation relations, i.e. it is a fermion, regardless of whether its expansion coefficients are eigenstates of any discrete symmetry operator or not. That is in complete agreement with the spin-statistics theorem. Finally, in the last section, we present our conclusions.


\section{General framework for spin-1/2 relativistic quantum fields} \label{sec2}

In this section, we establish the general conditions that any relativistic quantum field in $(3+1)$-dimensional spacetime must satisfy to represent a physical particle state. We do this by considering the fundamental principle of the Lorentz invariance of the $S$-matrix \cite{WEI}.  It requires that the interaction term is the spacetime integral of a Hamiltonian scalar density, ${\cal {H}}(x)$, that satisfies:
\begin{eqnarray}
U(\Lambda,\,a){\cal{H}}(x)U(\Lambda,a)^{-1}&=&{\cal{H}}(\Lambda x+a),\,
\end{eqnarray}
$\textrm{i.e. } { \cal{H}}(x) \textrm{ is a Lorentz scalar, and}\nonumber$
\begin{eqnarray}
\left[{\cal{H}}(x),{\cal{H}}(x')\right] &=&0,  \textrm{ for } (x-x')^2\geq 0, \label{hamiltonian}
\end{eqnarray}
where $\Lambda$ is a Lorentz transformation, $a$ is a general spacetime translation and, $U$ is the corresponding operator representing the Poincar\'e transformation in the Hilbert space. Also, note that throughout this work we use the metric signature $ (-, +, +, +) $.

In general, ${\cal{H}}(x)$ is constructed from the creation and annihilation operators to satisfy the cluster decomposition principle. Because of the transformation of these operators under the Lorentz group, ${\cal{H}}(x)$ must be built of fields $\Psi(x)$ (and its Hermitian conjugate) with the following form:
\br
\Psi_\ell (x) = \alpha\,\psi_\ell^+(x) + \beta\, \chi_\ell^{-}(x), \label{generalqf}
\er
where $\alpha$ and $\beta$ are constants which are chosen so that the condition in Eq.~\eqref{hamiltonian} is satisfied, and
\begin{eqnarray}
\psi_\ell^+(x) &=&\sum\limits_{s,n}\int d^3p\,\, u_\ell(x,\pv,s,n) \,{a}(\pv,s,n), \label{creation}\\
\chi_\ell^-(x) &=& \sum\limits_{s,n}\int d^3p\,\, v_\ell(x,\pv,s,n) \,b^\da(\pv,s,n),\label{annihilation}
\end{eqnarray}
where $n$ denotes internal quantum numbers, $s$ runs over the spin $z$-components and $\ell$ runs over the representation index. From here on, we will omit the $n$ index because it is not essential for our purposes.

The coefficients $u_\ell(x,\pv,s)$ and $ v_\ell(x,\pv,s)$ must be chosen so that the fields $\psi_\ell^+(x)$ and $\chi_\ell^-(x)$  transform under Lorentz transformations as
\begin{eqnarray}
U(\Lambda,a) \psi_\ell^+(x)U(\Lambda,a)^{-1} &=&\sum\limits_{\bar{\ell}} D_{\ell\bar{\ell}}(\Lambda^{-1})\psi_{\bar{\ell}}^{+}(\Lambda x+a), \label{creation1}\\
U(\Lambda,a) \chi_\ell^-(x) U(\Lambda,a)^{-1} &=&\sum\limits_{\bar{\ell}} D_{\ell \bar{\ell}}(\Lambda^{-1}) \chi_{\bar{\ell}}^-(\Lambda x+a), \label{annihilation1}
\end{eqnarray}
where the $D$-matrix furnishes a representation of the homogeneous Lorentz group \cite{Cornwell}.

More specific properties of the coefficients  $u_\ell(x,\pv,s)$ and $v_\ell(x,\pv,s)$ can be deduced by applying Eqs. \eqref{creation1} and \eqref{annihilation1} for pure translations, boosts and rotations. First, applying pure translations it is found \cite{WEI,Duncan}:
\begin{eqnarray}
u_\ell(x,\pv,s)&=&{(2\pi)^{-3/2}}\,e^{ip\cdot x}\,u_\ell(\pv,s),  \label{translations1}\\
v_\ell(x,\pv,s)&=&{(2\pi)^{-3/2}}\,e^{-ip\cdot x}v_\ell(\pv,s), \label{translations2}
\end{eqnarray}
where the factor ${(2\pi)^{-3/2}}$ is conventional. In other words, because of the translational invariance, the dependence on $x$ of $u_\ell(x,\pv,s)$ and $v_\ell(x,\pv,s)$ is only an exponential factor. We also have that due to the invariance under a general boost, $L(\pv)$, these coefficients must satisfy
\br
u_\ell(\mathbf{p}, s)&&=\sqrt{\frac{m}{p^{0}}} \sum\limits_{\bar{l}} \,D_{\ell \bar{\ell}}(L(\pv)) u_{\bar{\ell}}(0, s), \label{boost1} \\
v_{\ell}(\mathbf{p}, s)&&=\sqrt{\frac{m}{p^{0}}} \sum\limits_{\bar{\ell}} D_{\ell \bar{\ell}}(L(\pv)) v_{\bar{\ell}}(0,s), \label{boost2}
\er
where $D(L(\pv))$ is the matrix representation of a general Lo\-rentz boost and, $u(0,s)$ and $v(0,s)$ are the corresponding zero momentum coefficients.
Finally, we consider a general rotation $\Lambda=R$ in Eqs.~\eqref{creation1} and \eqref{annihilation1}. It is straightforward to see that the coefficients $u(0,s)$ and $v(0,s)$ have to satisfy the rotational symmetry conditions
\br
\sum_{\bar s} u_{\bar\ell}(0, \bar{s}) \mathbf{J}_{\bar{s} s}^{(j)}
&=&\sum_{\ell} {\cal J}_{\bar{\ell}\ell}\, u_{\ell}(0,s), \label{irrep1}\\
-\sum_{\bar{s}} v_{\bar{\ell}}(0, \bar{s}) \mathbf{J}_{\bar{s} s}^{(j)^{*}}
&=&  \sum_{\ell} {\cal J}_{\bar{\ell}\ell}\, v_{\ell}(0, s),\label{irrep2}
\er
where $\mathbf{J}^{(j)}$ and ${\cal J}$ are the angular momentum matrices in the representations $D^{(j)}(R)$ and $D(R)$, respectively \cite{Cornwell}. The conditions \eqref{irrep1} and \eqref{irrep2} mean that if $\psi^+(x)$ and $\chi^-(x)$ are supposed to describe particles with a spin $j$,
the spin-$j$ representation $D^{(j)}(R)$ must be among the irreducible components of the representation $D(R)$.

At this point, it is important to remark that the conditions on the fields $\psi_\ell^+(x)$ and $\chi_\ell^-(x)$ given in Eqs. \eqref{creation1}-\eqref{annihilation1} and their respective consequences in Eqs. \eqref{translations1}--\eqref{irrep2} are general in the sense that they do not depend on a particular adjoint structure or on a Lagrangian (equations of motion), or on the use of a discrete symmetry such a parity, time reversal or charge conjugation.

Then we apply the presented general framework to our specific case of spin-1/2 fields.
To do that, we use the Weyl basis to write $\mathbf{J}^{(j)}$ and ${\cal J}$ matrices in Eqs. \eqref{irrep1}--\eqref{irrep2} in the specific representations $j=1/2$ and $(\frac{1}{2},0)\oplus(0,\frac{1}{2})$, respectively. In this basis we have \cite{Georgi}
\br
 \mathbf{J}^{(1/2)} = \frac{1}{2}\boldsymbol{\sigma}, \qquad -\mathbf{J}^{(1/2)^{*}} = \frac{1}{2}\sigma_2\boldsymbol{\sigma}\sigma_2, \label{spinoperator}
\er
and
\br
{\cal J}_{i0} = -\frac{i}{2} \left[ \begin{array}{cc}
    \sigma_i &  {\mathbf 0}\\
    {\bf 0} & -\sigma_i
\end{array}\right], \qquad {\cal J}_{ij} = \frac{1}{2}\epsilon_{ijk} \left[ \begin{array}{cc}
    \sigma_k &  {\mathbf 0}\\
    {\bf 0} & \sigma_k
\end{array}\right], \label{spinoperator2}
\er
where $\sigma_k$, $k=1,2,3$ are the Pauli matrices. Substituting the above matrix representations into the conditions (\ref{irrep1}) and (\ref{irrep2}), we obtain
\br
\sum_{\bar s} (u_\pm (0, \bar{s}))_{i}\, \mathbf{J}_{\bar{s} s}^{(1/2)}
&=&\sum_{j} \frac{1}{2}\boldsymbol{\sigma}_{ij} (u_{\pm}(0,s))_j , \label{irrep3}\\
-\sum_{\bar{s}} (v_\pm(0, \bar{s}))_{i}\, \mathbf{J}_{\bar{s} s}^{(1/2)^{*}}
&=&\sum_{j} \frac{1}{2} \boldsymbol{\sigma}_{ij} (v_{\pm}(0, s))_j,\label{irrep4}
\er
where we have defined $u(0,s)\equiv(u_{+}(0,s), \,\,u_{ -}(0,s))^{\text T}$ and $v(0,s)\equiv(v_{+}(0,s), \,\,v_{ -}(0,s))^{\text T}$.
By considering $(u_\pm (0, {s}))_{{i}}$ and $(v_\pm (0, {s}))_{{i}}$ as the $(i,s)$ elements of corresponding matrices $U_\pm$ and $V_\pm$, we can rewrite the conditions (\ref{irrep1}) and (\ref{irrep2}) in matrix form
\br
U_\pm \, \mathbf{J}^{(1/2)} &=&  \frac{1}{2}\boldsymbol{\sigma} U_\pm ,\label{rot1}\\
-V_\pm \, \mathbf{J}^{(1/2)^*} &=&  \frac{1}{2} \boldsymbol{\sigma} V_\pm. \label{rot2}
\er
The last two expressions play a central role in the following sections, where we investigate the rotational invariance for the quantum fields.

\section{Spin-1/2 ``bosons" with mass dimension 3/2 case \label{sec3}}

We now turn our attention to the class of quantum fields proposed in Refs. \cite{dharamnewbosons,Ahluwalia:2021vfu,Ahluwalia:2019ujt}. In those references, it is stated that the spin-1/2 field belongs to the representation $(\frac{1}{2},0)\oplus(0,\frac{1}{2})$, but satisfies the Bose-Einstein statistics, which is in disagreement with the spin-statistics theorem. Therefore, in this section we will test whether the proposed fields satisfy the conditions derived from the Lorentz invariance in the previous section.

Specifically, the quantum field is defined as
\begin{equation}
\begin{split}
{\Phi}(x) =&\int \frac{d^3p}{(2\pi)^3}\, \frac{1}{\sqrt{2p_0}}\left[ \sum\limits_{i=1,2}\lambda_i(\pv) e^{-ip\cdot x} \,a_i(\pv)\right.\\
&+\left. \sum\limits_{i=3,4}\lambda_i(\pv) e^{+ip\cdot x} \,b_i^{\dagger}(\pv)\right],  \label{newfield}
\end{split}
\end{equation}
where the $\lambda_i(\pv)$ for any $\pv \neq 0$ is obtained by applying a Lorentz boost, $D(L(\pv))$, on the coefficients $\lambda_i(0)$ which are chosen to be linear combinations of the eigenstates $\mu_i,$ $i=1,\ldots,4$, of the matrix $i\gamma_{3}$, as follows
\begin{eqnarray}
\lambda_1(0)&&=\sqrt{\frac{m}{2}}\,(\mu_1+i\mu_2), \quad \lambda_2(0)=\sqrt{\frac{m}{2}}\,(\mu_1-i\mu_2),  \label{lambda1}\\
\lambda_3(0)&&=\sqrt{\frac{m}{2}}\,(\mu_3+i\mu_4), \quad \lambda_4(0)=\sqrt{\frac{m}{2}}\,(\mu_3-i\mu_4), \label{lambda3}
\end{eqnarray}
with
\br
\mu_1 &=& \left[\begin{array}{l} 0,\,i,\,0,\,1 \end{array} \right]^\textrm{T}, \qquad \quad \mu_2 \,=\,\left[\begin{array}{l} i,\,0,\,1,\,0\end{array} \right]^\textrm{T}, \label{mucoefficient1}\\[0.3cm]
\mu_3 &=& \left[\begin{array}{l} 0,\,-i,\,0,\,1 \end{array} \right]^{\textrm{T}}, \qquad \,
\mu_4 \,=\,\left[\begin{array}{l} -i,\,0,\,1,\,0 \end{array} \right]^\textrm{T}. \label{mucoefficient2}
\er

The key idea behind the above choice is to expand the coefficients of the quantum field as linear combinations of the eigenstates of one of the sixteen different roots of the identity matrix $\mathds {1} _ {4 \times4}$. Note that $\mathds {1} _ {4 \times4}$, $i\gamma_{1}$, $i\gamma_{2}$, $i\gamma_{3}$, $\gamma_{0}$, $i\gamma_{2}\gamma_{3}$, $i\gamma_{3}\gamma_{1}$, $i\gamma_{1}\gamma_{2}$, $\gamma_{0}\gamma_{1}$, $\gamma_{0}\gamma_{2}$, $\gamma_{0}\gamma_{3}$,  $i\gamma_{0}\gamma_{2}\gamma_{3}$, $i\gamma_{0}\gamma_{1}\gamma_{3}$, $i\gamma_{0}\gamma_{1}\gamma_{2}$, $\gamma_{1}\gamma_{2}\gamma_{3}$ and $i\gamma_{0}\gamma_{1}\gamma_{2}\gamma_{3}$ are all roots of the $\mathds {1} _ {4 \times4}$ matrix. From here on these roots are denoted as $\Omega_i$ with $i=1,\ldots,16$. So, the case in consideration corresponds to $\Omega_4$. Throughout this work we use the Weyl representation for the Dirac matrices
\br
\gamma_{0} = \left[ \begin{array}{cc}
    {\mathbf 0} &  \mathds{1}_{2\times 2} \\
     \mathds{1}_{2\times 2} & {\bf 0}
\end{array}\right], \qquad \gamma_{k} = \left[ \begin{array}{cc}
    {\mathbf 0} & \sigma_k \\
    -\sigma_k & {\bf 0}
\end{array}\right].
\er
It is also important to remark that in the momentum space the expansion coefficients of $\Phi(x)$,  $\lambda_i(\pv)$, satisfy Dirac-like equations as
\begin{eqnarray}
\left(a_\mu p^\mu- m\mathds{1}\right)\lambda_{1,2}(\pv)&=&0, \\
\left(a_\mu p^\mu+ m\mathds{1}\right)\lambda_{3,4}(\pv)&=&0,
\end{eqnarray}
where $a_\mu = i\gamma_5\gamma_\mu$, $\gamma_5=-\frac{i}{4!}\epsilon^{\mu\nu\lambda\sigma}\gamma_\mu\gamma_\nu\gamma_\lambda\gamma_\sigma$ with \mbox{$\epsilon^{0123}=1$} and, $a_\mu$ satisfies $\lbrace a_{\mu},a_{\nu}\rbrace=2\eta_{\mu\nu}\mathds{1}$. That means that the coefficients $\lambda_i(\pv)$ are eigenvectors of the $a_\mu p^\mu$ operator, and not of the parity operator ${\cal P}=m^{-1} p^{\mu}\gamma_\mu$.

Notice that the $\lambda_i(\pv)$ coefficients of the quantum field $\Phi(x)$ defined in Eq. \eqref{newfield} satisfy trivially the translational invariance condition Eqs. \eqref{translations1} and \eqref{translations2}, as well as  the invariance under a general boost Eqs. \eqref{boost1} and \eqref{boost2}. However, they do not satisfy the rotational symmetry conditions.  This can be seen by substituting in Eqs. \eqref{rot1} and \eqref{rot2}, the  $\mathbf{J}^{(1/2)}$representation Eq. \eqref{spinoperator} and, $U_\pm$ and $V_\pm$  matrices corresponding to $\lambda_i(0)$ coefficients, which in this case are
\br
U_{+} =-V_{+}=\sqrt{\frac{m}{2}} \left[ \begin{array}{cc}
    -1 &  1\\
     i & i
\end{array}\right];
U_{-} = V_{-}= \sqrt{\frac{m}{2}} \left[ \begin{array}{cc}
     i& -i\\
     1 & 1
\end{array}\right]. \label{umatrices}
\er
Also note that this result does not depend on the chosen labelling of the coefficients $\lambda_i(0)$ in Eqs. \eqref{lambda1} and \eqref{lambda3}. Even if the $\mu_i$ in Eqs. \eqref{mucoefficient1} and \eqref{mucoefficient2} were used as the expansion coefficients of the field ${\Phi}(x)$, the rotational symmetry condition would still not be satisfied.

A direct physical consequence of the lack of rotational symmetry invariance is that any quantum state created by $\Phi(x)$ does not have spin $1/2$. This can be seen by applying the spin operator in the specific representation $(\frac{1}{2},0)\oplus(0,\frac{1}{2})$ to a state created by the $\Phi(x)$ field. For instance, if the $\Phi(x)$ field in Eq. \eqref{newfield} is applied to the vacuum state, $\left.\vert 0 \right\rangle$, it creates a type-$b$ state, denoted as $\left.\vert 1_b \right\rangle$, which has two contributions coming from $\lambda_3$ and $\lambda_4$. Note that, since the spin is a static physical property, i.e. the spin and the  momentum operators commute, it is sufficient to work in center-of-momentum frame (where $\pv=0$) to determine the spin of such state. More specifically, by calculating the expectation value of any spin projection operator on  the $\left.\vert 1_b \right\rangle$ state, i.e. $\langle 1_b | \hat{S}_i\left.\vert 1_b \right\rangle$ with $i=x,y,z$, one obtains zero instead of $\hslash/2$, which it should be if the $\Phi(x)$ field created a physical state with spin $1/2$. The same conclusion is obtained when considering a type-$a$ state. Therefore, the field $\Phi(x)$ in Eq. \eqref{newfield} does not describe a physical particle state with spin $1/2$.

Another critical consequence of the absence of rotational symmetry is that the field defined as in Eq.~\eqref{newfield} is not covariant under the rotational group nor under the homogeneous Lorentz group, as it should be to satisfy Eq.~\eqref{creation1}. This implies that this field cannot be used to construct a rotationally invariant Hamiltonian density and therefore the angular momentum of a closed system is not conserved. This consequence is true even in a classical field theory and in a non-relativistic theory.

\section{Fermion with mass dimension 1 case \label{sec4}}

The quantum field proposed in Refs. \cite {mdobook,newfermions}, the so-called mass dimension 1 fermion, has the same form as given in Eq. \eqref{newfield} but in this case the $\lambda_i(0)$ coefficients are proportional to the eigenstates of $\Omega_7=i\gamma_3\gamma_1$, as follows
\begin{eqnarray}
\lambda_1(0)&&=\sqrt{m}\,[0,0,-i,1], \quad \lambda_2(0)=\sqrt{m}\,[0,0,i,1],  \label{lambda1x}\\
\lambda_3(0)&&=\sqrt{m}\,[-i,1,0,0], \quad \lambda_4(0)=\sqrt{m}\,[i,1,0,0]. \label{lambda3x}
\end{eqnarray}
Following the same procedure used in Sec. \ref{sec3}, it can be shown these $\lambda_i(0)$ do not satisfy the rotational symmetry conditions for spin $1/2$. In this case the $U_\pm$ and $V_\pm$  matrices are given as
\br
U_{+} =V_{-}= \left[ \begin{array}{cc}
     0 &  0\\
     0 & 0
\end{array}\right]; \quad
U_{-} = V_{+}= \sqrt{m} \left[ \begin{array}{cc}
     -i& i\\
     1 & 1
\end{array}\right]. \label{umatricesx}
\er
Furthermore, by doing the same calculation for the expectation value of the spinor operator, it can be seen that a state created by this field is not a physical state with spin $1/2$.

Analogously, we can consider the case of the ELKO quantum fields  \cite {Grumiller2005,PhysRevD.72.067701,mdobook,report}, which are constructed from the eigenstates of the charge conjugation operator ($\cal{C}$). In this case, we have the coefficients
\begin{equation}
\lambda_{\pm}^S\left(k^\mu\right)=\left[\begin{array}{c}
i \Theta\left[\phi_{\pm}\left(k^\mu\right)\right]^* \\
\phi_{\pm}\left(k^\mu\right)
\end{array}\right], \quad \lambda_{\pm}^A\left(k^\mu\right)=\left[\begin{array}{c}
\mp i \Theta\left[\phi_{\mp}\left(k^\mu\right)\right]^* \\
\pm \phi_{\mp}\left(k^\mu\right)
\end{array}\right],
\end{equation}
where ${\cal{C}}\lambda_{\pm}^S=+\lambda_{\pm}^S$, ${\cal{C}} \lambda_{\pm}^A=-\lambda_{\pm}^A$, $\Theta=-i\sigma_2$ and
\begin{equation}
\begin{aligned}
&\phi_{+}\left(k^\mu\right)=\sqrt{m}\left[\begin{array}{c}
\cos (\theta / 2) \exp (-i \varphi / 2) \\
\sin (\theta / 2) \exp (+i \varphi / 2)
\end{array}\right], \\
&\phi_{-}\left(k^\mu\right)=\sqrt{m}\left[\begin{array}{c}
-\sin (\theta / 2) \exp (-i \varphi / 2) \\
\cos (\theta / 2) \exp (+i \varphi / 2)
\end{array}\right],
\end{aligned}
\end{equation}
where $\theta$ and $\varphi$ are arbitrary angles that parametrize the components of $\phi_{+}\left(k^\mu\right)$ and $\phi_{-}\left(k^\mu\right)$. Note that $\lambda_{\pm}^S\left(k^\mu\right)$ and $\lambda_{\pm}^A\left(k^\mu\right)$ could be defined with some global phase factors ($\xi_1,\xi_2,\xi_3,\xi_4$). However, the specific value of these phases will not modify the following results. Thus, we  use $\xi_1=\xi_2,\xi_3=0$  and $\xi_4=\pi$ as in Ref. \cite{mdobook}.

The corresponding matrices $U_{\pm}$ and $V_{\pm}$ are
\begin{equation}
U_+=i\,V_{-}=\sqrt{m}\left[
\begin{array}{cc}
 -i  e^{-{i \varphi/2}} \sin \left(\theta/2\right) & -i e^{-i \varphi/2} \cos \left(\theta/2 \right) \\
 i  e^{i \varphi/2} \cos \left(\theta/2\right) & -i e^{i \varphi/2} \sin \left(\theta/2\right) \\
\end{array}
\right],
\end{equation}
and
\begin{equation}
U_{-}=-i\,V_{+}=\sqrt{m}\left[
\begin{array}{cc}
  e^{-i \varphi/2} \cos \left(\theta/2 \right) &  -e^{-i \varphi/2
   } \sin \left(\theta/2 \right) \\
  e^{i \varphi/2 } \sin \left(\theta/2 \right) & e^{i \varphi/2 }
   \cos \left(\theta/2\right)\\
\end{array}
\right].
\end{equation}
Substituting these matrices into Eqs. \eqref{rot1} and \eqref{rot2}, it is straightforward to see that the ELKO quantum fields are not invariant under rotations. Therefore, these fields do not represent a physical state with spin $1/2$ either.

Actually, there is a deeper reason why the spinors considered in both cases do not satisfy the rotational symmetry invariance. Since  $\mathbf{J}^{(1/2)}$, $-\mathbf{J}^{(1/2)^{*}}$ and the $\frac{1}{2}\mathbf{\sigma}$ are irreducible representations of the Lie algebra of the rotation group, the Schur's Lemma \cite{Georgi} can be applied to show that there exist only two possible solutions for the $U_\pm$ and $V_\pm$ matrices in Eqs. \eqref{rot1} and \eqref{rot2}. The first one is a trivial solution with vanishing matrices, and the second one tell us that $U_\pm$ and $V_\pm\s_2$ are proportional to the identity matrix. Thus, we find that the most general zero-momentum  $u_\ell(0,s)$ and $v_\ell(0,s)$ spinors can take only the following forms
\br
u\(0,{\frac{1}{2}}\) &=& \left[\begin{array}{l} c_+\\0\\c_-\\0
\end{array} \right], \qquad
u\(0,{-\frac{1}{2}}\) \,=\,\left[\begin{array}{l} 0\\c_+\\0\\c_-\end{array} \right], \label{eq2.12}\\[0.3cm]
v\(0,{\frac{1}{2}}\) &=& \left[\begin{array}{l} 0\\d_+\\0\\d_-
\end{array} \right], \qquad
v\(0,{-\frac{1}{2}}\) \,=\, -\left[\begin{array}{l} d_+\\0\\d_-\\0\end{array} \right],\label{eq2.13}
\er
where $c_\pm$ and $d_\pm$ are arbitrary constants, which in general can be complex numbers or even zero and, must satisfy $|c_{+}|^2$ $+|c_{-}|^2=1$ and $|d_{+}|^2+|d_{-}|^2=1$. Therefore, the expansion coefficients of all considered fields do not fit this particular form. Note that a quantum field, expanded by the coefficients in Eqs.~\eqref{eq2.12} and \eqref{eq2.13}, has the foreseen spin expectation value for a quantum field of spin $1/2$, i.e. $\langle 1_b | \hat{S}_i \vert 1_b \rangle$ $=\frac{\hslash}{2}$ with $i=x,y,z$.

We can now conclude the main reason why the spin-1/2 ``bosons'' with mass dimension $3/2$ and fermions with mass dimension 1, including ELKO, do not describe a physical particle state with spin $1/2$.
Although the constructions of those fields resemble Dirac's historic construction in some respects, they use a linear combination of the eigenstates of the $\Omega_4$, $\Omega_7$ and $\cal{C}$ operators as expansion coefficients of the quantum field. Therefore, those quantum fields are not covariant under rotations.
When the covariance of the quantum fields established in Eqs. \eqref{creation1} and \eqref{annihilation1} is applied to a pure rotation, that is, $\Lambda=R$ and $a=0$, we see that the quantum field, $\psi(x)$, has to transform as $U(R,0) \psi(x) U(R,0)^{-1} = D(R^{-1})\psi(R\,x)$, where $D(R^{-1})$ has to be a representation of the rotation group, that is, a representation of the well-known angular momentum. If we consider any of the proposed new fields as  $\psi(x)$, we obtain that $U(R,0) \psi(x) U(R,0)^{-1} \neq D(R^{-1})\psi(R\,x)$, where $D(R^{-1})$ is the $(\frac{1}{2},0) \oplus (0, \frac{1}{2})$ representation generated by the generators in Eq. \eqref{spinoperator2}.  Moreover, one could think that the rotational covariance is only broken by a global phase, i.e. $U(R,0) \psi(x) U(R,0)^{-1} = \exp(i\,\omega)\times D(R^{-1})\psi(R\,x)$, where $\omega$ is an arbitrary phase. However, this is not the case as can be seen by a straightforward calculation.

\section{Confronting spin-1/2 bosons with mass dimension 3/2 against the spin-statistics theorem \label{sec5}}

According to Refs. \cite{dharamnewbosons,Ahluwalia:2021vfu,Ahluwalia:2019ujt}, the so-called spin-1/2 ``bosons'' field $\Phi(x)$ and its adjoint commute, instead of anticommuting, for space-like separations. Consequently, it was concluded that $\Phi(x)$ is a new type of bosonic field endowed with spin $1/2$, which evades the well-stablished spin-statistics theorem. Furthermore, the Feynman-Dyson propagator was computed, showing that $\Phi(x)$ should have mass dimension $3/2$.

In order to elucidate this controversial result, we will prove that relativistic spin-1/2 fields, in $(3+1)$-dimensional spacetime, satisfy the Fermi-Dirac statistic regardless of whether its expansion coefficients are chosen to be eigenstates of the parity, charge conjugation, time reversal, or any other operator. To do that, we will generalize the Weinberg's proof \cite{WEI} without imposing any discrete symmetry.

Our starting point is the condition in Eq. \eqref{hamiltonian} that comes from the Lorentz invariance of the S-matrix, which ensures the inability of an observer to detect absolute inertial motion by performing scattering experiments. From an algebraic point of view, this condition implies that, for space-like intervals, the fields in \eqref{generalqf} satisfy the following conditions
\br
\left[\Psi_{\ell}(x), \Psi_{\bar\ell}(y)\right]_{\mp} &=& 0, \qquad \quad \left[\Psi_{\ell}(x), \Psi_{\bar\ell}^{\dagger}(y)\right]_{\mp} =  0,\label{eq2.14}
\er
where $(-,+)$ signs means commutator and anti-commutator, respectively. Note that the second condition in Eq. \eqref{eq2.14} makes use of the Hermitian conjugate field and then it is not necessary to define the adjoint structure at this point.

\noindent The first condition in Eq. (\ref{eq2.14}) is satisfied for any constants $\alpha$ and $\beta$ because
\br
&&\left[{a}(\pv,s), {a}(\pv',s')\right]_{\mp} = 0, \quad
 \left[{b}(\pv,s), {b}(\pv',s')\right]_{\mp} \,=\, 0, \nonumber \\
 && \left[{a}(\pv,s), {b}^\dagger(\pv',s')\right]_{\mp} \,= 0,\,
\er
and the corresponding relations for the Hermitian conjugate operators. However, the second condition in Eq. (\ref{eq2.14}) is not satisfied in general, which imposes some constraints on the type of statistics that the field satisfies and, eventually, on the $\alpha$ and $\beta$ constants.
Using the canonical relations
\br
\left[{a}(\pv,s), {a}^\dagger(\pv',s')\right]_{\mp} = \delta^{(3)}(\pv-\pv')\delta_{ss'},\nonumber \\
\left[{b}(\pv,s), {b}^\dagger(\pv',s')\right]_{\mp} = \delta^{(3)}(\pv-\pv')\delta_{ss'}, \label{canonicalrelation}
\er
we obtain
\begin{eqnarray}
\left[\Psi_{\ell}(x), \Psi_{\bar\ell}^{\dagger}(y)\right]_{\mp}&=&(2\pi)^{-3}\int d^{3} p\left[|\alpha|^{2} N_{\ell \bar{\ell}}(\mathbf{p}) e^{i p \cdot(x-y)}\right. \nonumber\\
&&\left. \mp|{\beta}|^{2} M_{\ell\bar{\ell}}(\mathbf{p}) e^{-i p \cdot(x-y)}  \right], \label{eq2.18}
\end{eqnarray}
where
\br
N_{\ell \bar{\ell}}(\mathbf{p}) &\equiv& \sum_{s} u_{\ell}(\mathbf{p}, s) u_{\bar\ell}^{*}(\mathbf{p}, s), \\
M_{\ell \bar{\ell}}(\mathbf{p}) &\equiv& \sum_{s} v_{\ell}(\mathbf{p}, s) v_{\bar{\ell}}^{*}(\mathbf{p}, s).
\er
Applying Eqs. \eqref{boost1} and \eqref{boost2}, we can write the above quantities as
\br
N_{\ell \bar{\ell}}(\mathbf{p}) &=& \frac{m}{p^0}D(L(\pv)) N_{\ell \bar{\ell}}(0) D^\dagger (L(\pv)), \label{eq2.21}\\
M_{\ell \bar{\ell}}(\mathbf{p}) &=& \frac{m}{p^0}D(L(\pv)) M_{\ell \bar{\ell}}(0) D^\dagger (L(\pv)),\label{eq2.22}
\er
where
\br
N_{\ell \bar{\ell}}(0)=\left[\begin{array}{cccc}
\left|c_{+}\right|^{2} & 0 & c_{+} c_{-}^{*} & 0 \\
0 & \left|c_{+}\right|^{2} & 0 & c_{+} c_{-}^{*} \\
c_{-}c_{+}^{*} & 0 & |c_{-}|^{2} & 0 \\
0 & c_{-}c_{+}^{*} & 0 & |c_{-}|^{2}
\end{array}\right], \label{Nx}
\er
and $M_{\ell{\bar\ell}}(0)$ is obtained by replacing $c_\pm \to d_\pm$ in Eq. \eqref{Nx}. Let us now consider the explicit calculations only for $N_{\ell \bar{\ell}}(0)$, which can be spanned in terms of the gamma matrices as
\begin{eqnarray}
\label{eq2.24}
N_{\ell \bar{\ell}}(0) &=& \frac{1}{2}\left(|c_+|^2 + |c_{-}|^2\right) \mathds{1} + \frac{1}{2} \left( c_+ c_{-}^* + c_{-}c_+^*\right) \gamma_0 \nonumber\\
&+&\frac{1}{2}\left(|c_+|^2 - |c_{-}|^2\right)\gamma_5+\frac{1}{2}\left(c_+c_{-}^* - c_{-}c_{+}^* \right)\gamma_0\gamma_5.
\end{eqnarray}
It is also worth to reinforce that Eq.~(\ref{eq2.24}) is a general result, where no assumptions on parity or any other discrete symmetries have been used.
Now, substituting Eq.~(\ref{eq2.24}) into Eq.~(\ref{eq2.21}) we obtain
\begin{eqnarray}
\begin{aligned}
\label{eq2.29}
N_{\ell \bar{\ell}}(\mathbf{p}) = \frac{1}{2p^0} &\Big[-{\left(|c_+|^2 + |c_{-}|^2\right)}{p^\mu \gamma_\mu}\\
 &+{m\left( c_+ c_{-}^* + c_{-}c_+^*\right)}\mathds{1}\\
&+{m \left(c_+c_{-}^* - c_{-}c_{+}^*\right)}\gamma_5 \\
& -\left(|c_+|^2 - |c_{-}|^2\right){p^{\mu} \gamma_\mu}\gamma_5 \Big]  \g_0.
\end{aligned}
\end{eqnarray}
With these results we can finally write Eq.~(\ref{eq2.18}) as
\br
\begin{aligned}
&\left[\Psi_{\ell}(x), \Psi_{\bar\ell}^{\dagger}(y)\right]_{\mp} = \Big( |\alpha|^2 \Big[{\left(|c_+|^2 +|c_{-}|^2\right)}{i\gamma_\mu\pa^\mu} \\
&\phantom{aaa} +{m\left( c_+ c_{-}^* + c_{-}c_+^*\right)}\mathds{1}+ {m \left(c_+c_{-}^* - c_{-}c_{+}^*\right)}\gamma_5   \\
&\phantom{aaa} +\left(|c_+|^2 - |c_{-}|^2\right) i\gamma_\mu \gamma_5 \pa^\mu \Big]  \g_0 \,\Delta_{+}(x-y) \hspace{2.0cm}\\
&\phantom{aaa} \mp  |\beta|^2 \Big[{\left(|d_+|^2 + |d_{-}|^2\right)}{i\gamma_\mu\pa^\mu} \\
&\phantom{aaa} +{m\left( d_+ d_{-}^* + d_{-}d_+^*\right)}\mathds{1} + {m \left(d_+d_{-}^* - d_{-}d_{+}^*\right)}\gamma_5\\
&\phantom{aaa} +\left(|d_+|^2 - |d_{-}|^2\right) i\gamma_\mu \gamma_5 \pa^\mu \Big] \g_0\, \Delta_{+}(y-x) \Big)_{\ell{\bar\ell}}\,\,, \qquad \quad \mbox{}
\end{aligned}
\er
where
\br
\Delta_{+}(x) \equiv \int \frac{d^3p}{2p^0(2\pi)^3} \,e^{ip\cdot x}
\er
is the Wightman 2-point function and $p^0=\sqrt{\mathbf{p}^2+m^2}$.
In order to satisfy the second condition in Eq.~\eqref{eq2.14}, it is necessary and sufficient that
\br
|\alpha|^2 |c_+|^2  = \mp |\beta|^2 |d_+|^2 , \quad  |\alpha|^2 |c_-|^2  = \mp |\beta|^2 |d_-|^2, \label{eq2.33}
\er
and
\br
|\alpha|^2  c_+ c_{-}^*  = \pm |\beta|^2  d_+ d_{-}^*, \label{eq2.34}
\er
where we have used that for $(x-y)$ space-like $\Delta_{+}(x-y)$ and its first derivative are even and odd functions of \mbox{$(x-y)$}, respectively. We notice that the constraints in Eq. \eqref{eq2.33} have a non-trivial solution provided we choose the bottom sign. This means that the spin-1/2 field $\Psi(x)$ in the $(\frac{1}{2},0)\oplus(0,\frac{1}{2})$ representation must satisfy the anti-commutation relations, i.e. the spin-1/2 field $\Psi(x)$ is a fermion, in complete agreement with the spin-statistics theorem. This general result does not depend on the use of any discrete symmetry such as parity or the definition of any adjoint structure.

Once the statistics of the $\Psi(x)$ field is set, Eqs. \eqref{eq2.33} and \eqref{eq2.34} simplify a little more and we obtain the following useful relation
\br
\frac{c_+}{c_-} = - \frac{d_+}{d_-}.\label{eq2.36}
\er
Note that $c_\pm$ and $d_\pm$ are not completely determined from the condition in Eq. \eqref{eq2.36}. Now, we can write $\alpha=|\alpha| e^{i\th_\alpha}$, $\beta=|\beta| e^{i\th_\beta}$ and, redefining the creation and annihilation operators in Eqs. (\ref{creation}) and (\ref{annihilation}) as $a(\pv,s)\to e^{i\th_\alpha}a(\pv,s)$ and $b(\pv,s)\to e^{-i\th_\beta}b(\pv,s)$ (note that these redefinitions do not modify the canonical anti-commutation relations \eqref{canonicalrelation} of $a$ and $b$ operators), we obtain
\br
\Psi(x) = |\alpha|\left[ \psi^+(x) + \frac{|\beta|}{|\alpha|}\chi^- (x)     \right].
\er
Absorbing the overall factor $|\alpha|$ in the normalization of the field $\Psi(x)$ and using the constraints in Eqs. (\ref{eq2.33}) and (\ref{eq2.34}), we get
\br
\Psi(x)= \psi^+(x) + \frac{|c_{\pm}|}{|d_{\pm}|}\chi^-(x). \label{finalfield}
\er
Needless to say, we are assuming that $|d_+|\neq 0$ or $|d_-|\neq 0$ in Eq. \eqref{finalfield}. In order to determine $c_\pm $ and $d_\pm$  completely, additional physical or mathematical conditions must be imposed. For example, a default choice is to impose parity symmetry on the fields in Eqs. \eqref{creation} and \eqref {annihilation} which lead us to
\br
\left(\frac{c_+}{c_-}\right)^2 = \left( \frac{d_+}{d_-}\right)^2=1. \label{eq2.37}
\er
Using this relation and the freedom of the overall factor, the form of the $\Psi(x)$ field is the standard Dirac field, i.e. $\Psi(x)= \psi^+(x) + \chi^- (x)$. It is also possible to consider the \linebreak charge-conjugation and time reversal properties of the field in Eq. \eqref{finalfield} to obtain, for instance, the Majorana fields \cite{WEI}.

\section{Conclusions} \label{conclusions}

In this paper, we show that although the so-called spin-1/2 ``bosons'' with the mass dimension $3/2$ and fermions with mass dimension 1 (including ELKO) satisfy some of the conditions arising from Lorentz invariance, such as translation, Eqs. \eqref{translations1} and \eqref{translations2}, and boost invariance, Eqs. \eqref{boost1} and \eqref{boost2}, they do not satisfy the invariance under rotations, Eqs. \eqref{irrep1} and \eqref{irrep2}. This brings unacceptable physical consequences considering a relativistic quantum theory. For example, none of the proposed fields create particles with definite spin. Furthermore, it is impossible to construct a rotationally invariant quantity from those fields, which implies that angular momentum is not conserved. In conclusion, such fields do not represent physical particle states with spin $1/2$.

Next, we prove that any spinorial field in the representation $(\frac{1}{2}, 0) \oplus (0, \frac{1}{2})$ of the Lorentz group satisfies the Fermi-Dirac statistics, that is, it is a fermionic field, regardless of whether its expansion coefficients are eigenstates of the parity operator or any other discrete symmetry operator. No particular definition of adjoint structure is used. In this way, we clarify that the reason why the so-called spin-1/2 ``boson'' escapes the spin-statistics theorem is that this field is not rotationally invariant and therefore not even a field with definite spin $1/2$.



\begin{acknowledgements}
\noindent
{A. R. Aguirre thanks CAPES and B. A. Couto e Silva thanks FAPEMIG for financial support. B. L. S\'anchez-Vega thanks the National Council for Scientific and Technological Development of Brazil, CNPq, for the financial support through grant n$^{\circ}$ 311699/2020-0. M. M.  Chaichian is deeply grateful to Dharam Ahluwalia for correspondence with several critical and clarifying remarks, to Christian B\"ohmer, Michael D\"utsch and Markku Oksanen for many useful discussions, enlightening suggestions and encouragement.} \end{acknowledgements}


\begin{thebibliography}{99}


\bibitem{WIG}
E. P. Wigner,
{\it On unitary representations of the inhomogeneous Lorentz group},
Annals Math. {\bf 40}, 149 (1939).

\bibitem{WIG2}
E. P. Wigner,
{\it Unitary representations of the inhomogeneous Lorentz group including reflections},
in Group theoretical concepts and methods in elementary particle physics, Lectures of the Istanbul summer school of theoretical physics, 37-80 (1964).

\bibitem{weinbergfeynman}
S. Weinberg,
{\it Feynman rules for any spin},
Phys. Rev. {\bf 133}, B1318 (1964).

\bibitem{wic}
E. H. Wichmann and J. H. Crichton,
{\it Cluster decomposition properties of the S matrix}.
Phys. Rev. {\bf 132}, 2788 (1963).

\bibitem{WEI}
S. Weinberg,
{\it The Quantum Theory of Fields, Vol. I: Foundations},
Cambridge University Press, New York, (1996).

\bibitem{Duncan}
A. Duncan,
{\it The Conceptual Framework of  Quantum Field Theory},
Oxford University Press, Oxford, (2012).



\bibitem{Grumiller2005}
D. V. Ahluwalia-Khalilova and D. Grumiller,
{\it Spin-half fermions with mass dimension 1: theory, phenomenology, and dark matter},
JCAP {\bf 07}, 012 (2005)

\bibitem{PhysRevD.72.067701}
D. V. Ahluwalia-Khalilova and D. Grumiller,
{\it Dark matter: A spin-1/2 fermion field with mass dimension 1?},
Phys. Rev. D {\bf 72}, 067701 (2005).


\bibitem{lounesto}
P. Lounesto,
{\it Clifford Algebras and Spinors, 2nd ed., London Mathematical Society Lecture Note Series},
Cambridge University Press, (2001).

\bibitem{mdobook}
D. V. Ahluwalia,
{\it Mass dimension one fermions},
Cambridge University Press (2019);

\bibitem{newfermions}
D. V. Ahluwalia,
{\it A new class of mass dimension 1 fermions,}
Proceedings of the Royal Society A {\bf 476}, 2240 (2020);

\bibitem{Boehmer1}
C. G. B{\"o}hmer,
{\it The Einstein-Cartan-Elko system},
Annalen Phys. {\bf 16}, 38 (2007).

\bibitem{Boehmer2}
C. G. B{\"o}hmer,
{\it The Einstein-Elko system - Can dark matter drive inflation?},
Annalen Phys. {\bf 16}, 325 (2007).

\bibitem{Boehmer3}
C. G. B{\"o}hmer,
{\it Dark spinor inflation - theory primer and dynamics},
Phys. Rev. D {\bf 77}, 123535 (2008).

\bibitem{Boehmer4}
C. G. B{\"o}hmer and J. Burnett,
{\it Dark spinors with torsion in cosmology},
Phys. Rev. D {\bf 78}, 104001 (2008).

\bibitem{Boehmer5}
C. G. B{\"o}hmer, J. Burnett, D. F. Mota and D. J. Shaw,
{\it Dark spinor models in gravitation and cosmology},
JHEP {\bf 07}, 053 (2010).

\bibitem{Fabbri}
L. Fabbri,
{\it The most general cosmological dynamics for ELKO matter fields},
Phys. Lett. B {\bf 704}, 255 (2011).

\bibitem{Sadjadi}
H. M. Sadjadi,
{\it On coincidence problem in ELKO dark energy model},
Gen. Relativ. Gravit. {\bf 44}, 2329 (2012).

\bibitem{Basak}
A. Basak, J. R. Bhatt, S. Shankaranarayanan and K. V. P. Varma,
{\it Attractor behaviour in ELKO cosmology},
JCAP {\bf 04}, 025 (2013).

\bibitem{daRocha:2007pz}
R. da Rocha and J. M. Hoff da Silva,
{\it From Dirac spinor fields to ELKO},
J .  Math.  Phys. {\bf 48}, 123517 (2007).

\bibitem{daRocha:2011yr}
R. da Rocha, A. E. Bernardini and J. M. Hoff da Silva,
{\it Exotic Dark Spinor Fields},
JHEP {\bf 1104}, 110 (2011).

\bibitem{Fabbri2}
L. Fabbri and S. Vignolo,
{\it ELKO and Dirac Spinors seen from Torsion},
Int. J. Mod. Phys. D {\bf 23}, 1444001 (2014).


\bibitem{Agarwal:2014oaa}
 B. Agarwal, P. Jain, S. Mitra, A. C. Nayak and R. K. Verma,
 {\it ELKO fermions as dark matter candidates},
 Phys.\ Rev.\ D {\bf 92}, 075027 (2015).


\bibitem{Alves:2014kta}
A. Alves, F. de Campos, M. Dias and J. M. Hoff da Silva,
{\it Searching for Elko dark matter spinors at the CERN LHC},
 Int. J. Mod. Phys. A {\bf 30}, 1550006 (2015).

\bibitem{HoffdaSilva:2016ffx}
J. M. Hoff da Silva, C. H. Coronado Villalobos, R. J. Bueno Rogerio and E. Scatena,
{\it On the bilinear covariants associated to mass dimension 1 spinors},
Eur.\ Phys.\ J.\ C {\bf 76}, 563 (2016).


\bibitem{CYL}
C.-Y. Lee and M. Dias,
{\it Constraints on mass dimension 1 fermionic dark matter from the Yukawa interaction},
Phys. Rev. D {\bf 94}, 065020 (2016).

\bibitem{nogo}
D. V. Ahluwalia,
{\it Evading Weinberg's no-go theorem to construct mass dimension 1 fermions: constructing darkness},
EPL {\bf 118}, 60001 (2017).

\bibitem{ahluwa1}
D. V. Ahluwalia,
{\it The theory of local mass dimension 1 fermions of spin one half},
Adv. Appl. Clifford Algebras {\bf 27}, 2247 (2017).

\bibitem{Rogerio2017}
R. J. Bueno Rogerio, J. M. Hoff da Silva, M. Dias and S. H. Pereira,
{\it Effective lagrangian for a mass dimension 1 fermionic field in curved spacetime},
JHEP {\bf 1802}, 145 (2018).

\bibitem{elkostates}
J. M. Hoff da Silva, and R. J. Bueno Rogerio,
{\it Massive spin-one-half one-particle states for the mass-dimension-one fermions},
EPL {\bf 128}, 11002 (2019).

\bibitem{report}
D. V. Ahluwalia, J. M. Hoff da Silva, C. Y. Lee, Y. X. Liu, S. H. Pereira,
et al. {\it Mass dimension one fermions: Constructing darkness},
Phys. Rept. 967 (2022).

\bibitem{dharamnewbosons}
D. V. Ahluwalia,
{\it Spin-half bosons with mass dimension three-half: Towards a resolution of the cosmological constant problem},
EPL {\bf 131}, 41001 (2020).

\bibitem{Ahluwalia:2021vfu}
D.~V.~Ahluwalia,
{\it New species of fermions and bosons, cosmological constant problem and a farewell to spin\textendash{}statistics theorem},
Int. J. Mod. Phys. D \textbf{30}, no.14, 2142031 (2021).

\bibitem{Ahluwalia:2019ujt}
D. V. Ahluwalia,
{\it Theory of spin one half bosons},
[arXiv:1908.09627 [physics.gen-ph]].

\bibitem{Curceanu}
C. Curceanu, J. D. Gillaspy, and R. C. Hilborn,
{\it Resource Letter SS–1: The Spin-Statistics Connection}
Am. J. Phys. {\bf 80}, 561 (2012).

\bibitem{Pauli1936}
W. Pauli,
{\it Th\'eorie quantique relativiste des particules obeisant \'a la statistique de Einstein–Bose.}
Annals de Institut Henry Poincar\'e, {\bf 6}, 109–136  (1936).

\bibitem{Massimi}
M. Massimi and M. Redhead,
{\it Weinberg's proof of the spin-statistics theorem},
Studies in History and Philosophy of Science Part B: Studies in History and Philosophy of Modern Physics, {\bf 34}, 4 (2003).

\bibitem{Pauli1940}
W. Pauli,
{\it The connection between spin and statistics}.
Physical Review, {\bf 58}, 716–722 (1940).

\bibitem{Wightman}
R. F. Streater and A. S. Wightman, {\it ”PCT, Spin and Statistics, and All That”}, (New York, W.A. Benjamin, 1964).

\bibitem{Goldhaber}
M. Goldhaber and Gertrude Scharff-Goldhaber.
{\it Identification of Beta-Rays with Atomic Electrons.}
Physical Review, {\bf 73}(12), 1472–1473 (1948).

\bibitem{Reines1974954}
F. Reines and H. W. Sobel,
{\it Test of the pauli exclusion principle for atomic electrons}
Phys. Rev. Lett. {\bf 32}, 954 (1974).

\bibitem{Amado19801338}
R. D. Amado and H. Primakoff,
{\it Comments on testing the Pauli principle }
Phys. Rev. C {\bf 22}, 1338 (1980)

\bibitem{Deilamian19954787}
K. Deilamian and J. D. Gillaspy and D. E. Kelleher,
{\it Search for small violations of the symmetrization postulate in an excited state of helium}
Phys. Rev. Lett. {\bf 74}, 4787 (1995).

\bibitem{Borexino:2004hfc}
H. O. Back and et. al.
{\it New experimental limits on violations of the Pauli exclusion principle obtained with the Borexino counting test facility}
EPJC {\bf 37} 421 (2004)


\bibitem{sym14050893}
F. Napolitano, S. Bartalucci, S. Bertolucci, M. Bazzi and et. al.
{\it Testing the Pauli Exclusion Principle with the VIP2 Experiment}.
Symmetry, {\bf 14}, 893 (2022).

\bibitem{PhysRevLett.104.253604}
D. English and V. V. Yashchuk and D. Budker.
{\it Spectroscopic Test of Bose-Einstein Statistics for Photons}
Phys. Rev. Lett. {\bf 104}, 253604 (2010).

\bibitem{Cornwell}
J. F. Cornwell,
{\it Group Theory in Physics Techniques of Physics, Vol. 2},
Academic Press, (1986).

\bibitem{Georgi}
Howard Georgi,
{\it Lie Algebras in Particle Physics},
Perseus Books, (1999).

\bibitem{Humphreys}
J. Humphreys,
{\it Introduction to Lie Algebras and Representation Theory, Graduate Texts in Mathematics 9},
Springer, (1972).

\end{thebibliography}
\end{document}